\documentstyle[epsf,preprint,aps]{revtex}
\begin{document}
\draft
\voffset=-30pt
\preprint{\it Physics Letters\/ \bf A 223\rm, 241-245 (1996) 
\hbox to 9cm{\hfill}}
\title{Resonance Energy--Exchange--Free Detection and\\  
`Welcher Weg' Experiment}
\author{Mladen Pavi\v ci\'c}
\address{Max--Planck--AG Nichtklassische Strahlung,  
Humboldt University of Berlin, D--12484 Berlin, Germany;\\
Department of Mathematics, University of Zagreb, GF, Ka\v ci\'ceva,  
POB 217, HR--10001 Zagreb, Croatia; E--mail: mpavicic@faust.irb.hr; 
Web page: http://m3k.grad.hr/pavicic} 

\maketitle \widetext \begin{abstract} 
It is shown that a monolithic total--internal--reflection resonator 
can be used for energy--exchange--free detections of objects without 
recoils. Related energy--exchange--free detection of \it `welcher 
Weg'\/ \rm is discussed and an experiment with an atom interferometer 
is proposed. The obtained results are in agreement with quantum theory. 
\end{abstract} 

\bigskip 

\narrowtext \section{Introduction} \label{sec:intro} 
Recently the old quantum \it welcher Weg\/ \rm(\it which--path\/\rm) 
reasoning has been used to devise experiments in which there is a 
certain probability of detecting an object without transferring a 
single quantum of energy to it. \cite{bomb,vaid94,z95,zz95,hans,P&I} 
The experiments are usually called \it interaction--free\/ \rm 
experiments but we use the name \it energy--exchange--free\/ \rm 
experiments in order to stress the fact that the detected object 
do interact with the measuring apparatus even when no quantum of 
energy $h\nu$ is transferred to 
it.\footnote{\parindent=0pt\fontsize{10pt}{11pt}
\selectfont\rm A slight twist (in brackets) of Niels Bohr's words 
might illuminate our decision: \it``It is true that in the 
measurements under consideration any direct mechanical interaction 
of the system and the measuring agencies is excluded, but \dots  
the procedure of measurements has an essential influence on the 
conditions on which the very definition of the physical quantities 
in question rests\dots [T]hese 
conditions must be considered as an inherent element of any 
phenomenon to which the term ``[interaction]'' can be 
unambiguously applied.''\/~\rm \protect\cite{bohr}} In effect, the 
reasoning, in an ideal case, is the following one. After the second 
beam splitter of a Mach--Zehnder interferometer one can always put 
a detector in such a position that it will never (i.e., with 
probability zero) detect a photon. If it does, then we are certain 
that an object blocked the \it ``other''\/ \rm path of the 
interferometer. Mach--Zehnder interferometer itself cannot be used 
for practical energy--exchange--free measurement because of its very low
efficiency (under 30\%). Therefore Paul and Pavi\v ci\'c \cite{P&I} 
recently proposed a very simple and easily feasible energy--exchange--free 
experiment based on the resonance in a single cavity which efficiency 
can realistically reach 95\%. As a resonator the proposal used a 
coated crystal which however reduced its efficiency.  

In this paper (in Sec.~\ref{sec:heisenberg}) we use a monolithic 
total--internal--reflection resonator which has recently shown 
extremely high efficiencies in order to construct an optical 
energy--exchange--free device with an efficiency approaching 100\%.  
Since the device differs from the usual quantum measurement devices, 
which assume an exchange of at least one quantum of energy \cite{hans}, 
it immediately provokes a question whether one carry out a \it 
welcher Weg\/ \rm interference experiment with its help. 
In Sec.~\ref{sec:path} we propose such an experiment using atom 
interferometry.  

\section{Resonance Energy--Exchange--Free Detection}
\label{sec:heisenberg}

The experiment (see Fig.~\ref{exp}) uses an uncoated monolithic 
total--internal--reflection resonator (MOTIRR) coupled to two 
triangular prisms by the frustrated total internal reflection 
(FTIR). \cite{motirr1,motirr2}. Both MOTIRR and prisms require a 
refractive index $n>1.41$ to achieve total reflection. 
When we bring prisms within a distance of the order of the 
wavelength, the total reflection within the resonator will be \it 
frustrated\/ \rm and a fraction of the beam will tunnel out of and 
into the resonator. Depending on the dimension of the gap and the 
polarization of the incidence beam one can well define reflectivity 
$R$ within the range from $10^{-5}$ to 0.99995. \cite{motirr2,ftir} 
Losses for MOTIRR and FTIR may be less than 0.3\%. 
The incident laser beam is chosen to be polarized perpendicularly to 
the incident plane so as to give a unique reflectivity for each 
photon. The faces of the resonator are polished spherically to give 
a large focusing factor and to narrow down the beam. A 
cavity which the beam in its round--trips has to go through is cut in 
the resonator and filled with an index--matching fluid to reduce losses. 
If there is an object in the cavity, i.e., in the way of the 
round--trips of the beam in the resonator, the incident 
beam will be almost totally reflected (into $D_r$). If there is 
no object, the beam will be almost totally transmitted (into $D_t$).      
As a source of the incoming beam a continuous wave laser (e.g., Nd:YAG) 
should be used because of its coherence length (up to 300$\>$km) and of 
its excellent frequency stability (down to 10$\>$kHz in the visible 
range). \cite{laser} 

We calculate the intensity of the beam arriving at detector $D_r$ when 
there is no object in the cavity in the following way. 
The portion of the incoming beam of amplitude $A(\omega)$ reflected at
the incoming surface is described by the amplitude 
$B_0(\omega)=-A(\omega)\sqrt{R}$,
where $R$ is reflectivity. The remaining part of the beam tunnels 
into MOTIRR and travel around guided by one FTIR (at the face 
next to the right prism where a part of the beam tunnels out into 
$D_t$) and by two proper total internal reflections. 
After a full round--trip the following portion of this beam joins the 
directly reflected portion of the beam by tunnelling into the left prism:
$B_1(\omega)=A(\omega)\sqrt{1-R}\sqrt{R}\sqrt{1-R}\>e^{i\psi}$, where 
$\psi=(\omega-\omega_{res})T$ is the phase added by each round--trip; 
here $\omega$ is the frequency of the incoming beam, $T$ is the 
round--trip time, and $\omega_{res}$ is the selection frequency 
corresponding to a wavelength which satisfies $\lambda=L/k$, where $L$ 
is the round--trip length of the resonator and $k$ is an integer.
Each subsequent round--trip contributes to a geometric progression
\begin{eqnarray}
B(\omega)=\sum^n_{i=0}B_i(\omega)\,,
\label{eq:total} 
\end{eqnarray}
where $n$ is the number of round--trips. We lock the laser at $\omega$ 
which is as close to $\omega_{res}$ as possible. Because of the 
afore--mentioned characteristics of the continuous wave lasers 
we can describe the input beam coming from such a laser during the 
coherence time by means of $A(\omega)=A\delta(\omega-\omega_{res})$. 
The following ratio of intensities of the reflected and the incoming 
beam then describes the efficiency of the device for free round--trips: 
\begin{eqnarray}
\eta={\int^\infty_0B(\omega)B^*(\omega)d\omega\over 
\int^\infty_0A(\omega)A^*(\omega)d\omega}=
1-{1-R\over1+R}[R^{2n}-1+2\sum^n_{j=1}(1+R^{2n-2j+1})R^{j-1}]\,.
\label{eq:sum}
\end{eqnarray}
The expression is obtained by mathematical induction from 
the geometric progression of the amplitudes [Eq.(\ref{eq:total})].

In the experiment one has to lower the intensity 
of the beam until it is likely that only one photon would appear 
within an appropriate time window (1$\>$ns---1$\>$ms $<$ coherence 
time) what allows the intensity in the cavity to build up. The 
obtained $\eta$ thus becomes a probability of detector
$D_r$ reacting when there is no object in the system. As shown in 
Fig.~\ref{eff}, $\eta$ approaches zero after 100 round--trips for $R=0.95$, 
after 1000 round--trips for $R=0.995$, etc., which is all  
multiply assured by continuous wave laser coherence length. In other 
words, a response from $D_r$ means that there is an object in the 
system. In the latter case the probability of the response is $R$, the 
probability of a photon hitting the object is $R(1-R)$, and the 
probability of photon exiting into $D_t$ detector is $(1-R)^2$. 
By widening the gaps between the resonator and the prisms 
we can make $R\rightarrow 1$ and therewith obtain an arbitrarily low 
probability of a photon hitting an object. We start each testing by 
recording the first two or three clicks of $D_r$ or $D_t$ after 
opening a gate for the incident beam. In this way we allow the 
beam to `wind up' in MOTIRR. And when either $D_r$ or $D_t$ fires 
(possibly even two or three times in a row to be sure in the 
result) the testing is over. Waiting for several clicks results in a 
bigger time window, but a chance of a photon hitting an object
remains 
very low. A possible 300$\>$km coherence length does not leave any 
doubt that a real experiment of detecting objects without 
transferring a single quantum of energy to them can be carried out 
successfully, i.e., with an efficiency exceeding 99\%. 
Also detectors might fail to react but this is not a
problem because single photon detectors with 85\%\ efficiency are 
already available and this would again only increase the 
time window for a few nano seconds what does not significantly 
influence the result. 

Thus we obtain the energy--exchange--free detection device 
in which the observed particles do not suffer any recoil. 
With opaque particles bigger than the wavelength 
of the applied laser beam we have got the maximal efficiency. 
However, our device can also see smaller 
objects because the main process in our resonator (which is a kind of 
the Fabry--Perot interferometer) is an interference in which the main 
role plays a possibility (which need not be realized) of a photon 
to hit an object in one of the round trips inside MOTIRR. 
In other words the device ``sees'' objects 
which exceed the resolution power of a standard microscope. The 
efficiency $1-\eta$ continuously decreases for smaller and smaller  
objects but that can be significantly improved if we choose the 
laser beam frequency which would correspond to an atomic 
resonance frequency of the object. On the other hand, the efficiency 
would be increased by using plasma X--ray lasers, if one designed an 
efficent X--ray resonator. For example, Nd$^{3+}$:glass laser system 
at Lawrence Livermore National Laboratory produces 250--ps X--ray 
laser pulses at wavelengths shorter than 5$\>$nm. \cite{teich} 
Our elaboration in Paul and Pavi\v ci\'c \cite{P&I} shows that    
the resonator would work with 250--ps pulses and the geometrical 
round path of 4$\>$cm.                                                                            
\narrowtext \section{`Welcher Weg' Detection} \label{sec:path} 

The experiment (see Fig.~\ref{path}) uses a combination of atom 
interferometer with ultracold metastable atoms and the resonance 
energy--exchange--free path detection by means of a movable MOTIRR 
(of course, without liquid what only slightly reduces the 
efficiency). To increase the probability of an
atom being hit by the round tripping beam, the incoming laser beam 
should be split into many beams by multiple beam splitters, each 
beam containing in average one photon in the chosen time window, 
so as to feed MOTIRR through many optical fibers. As for atom 
interferometer we adapt the one presented by Shimizu et al. 
\cite{cool} primarily because their method is almost background 
free. The atom source is a magneto--optical 
trap containing 1s$_5$ neon metastable atoms which are then 
excited to the 2p$_5$ state by a 598--nm laser beam. Of all the 
states to which 2p$_5$ decays we follow only 1s$_3$ atoms whose 
trajectory are determined only by the initial velocity and gravity 
(free fall from the trap). (Other states are either trapped by 
the magnetic field of the trap, or influenced and dispersed by 
another 640--nm cooling laser beam.) Now the atoms fall with 
different velocities but each velocity group forms interference 
fringes calculated as for the optical case and only corrected 
by a factor which arises from the acceleration by the gravity 
during the fall. MOTIRR is mounted on a device which follows 
(with acceleration) one velocity group from the double slit to 
microchannel plate detector (MCP). (Atoms from other groups 
move with respect to MOTIRR and therefore---because of their 
small cross section---cannot decohere MOTIRR). The laser is tuned 
to a frequency equal to the 1s$_3$ resonance frequency. The most 
distinguished fringes has the group which needs 0.1$\>$s to 
reach MCP from the double slit and are accelerated to 2$\>$m/s. 
The source is attenuated so much that there is in average only 
one atom in a velocity group. The whole process repeats 
every 0.4$\>$s.Assuming that we have 10$\>$ns recovery time 
for the photon detectors and 300 optical fibers we arrive at about 
10$^7$ counts which all go into one detector D$_t$ when no atom 
obstructs a round trip. (For reflectivity $R=0.999$ the probability 
of D$_r$ being activated is $2\cdot10^{-9}$.)  As soon as D$_r$ 
detector fires we know which slit the observed atom passed through. 
(The probability of photon hitting an atom is 0.001. In order to 
be able to estimate how many photons fired D$_r$ we can use 
\it photon chopping\/ \rm developed by Paul, T\"orm\"a, Kiss, and 
Jex \cite{chop}.) After $10^3$ repeating of 
such successful detections we have enough data to see whether the 
interference fringes are destroyed significantly with respect to 
unmonitored reference samples or not.    

\narrowtext \section{Discussion} \label{sec:con} 

In Sec.~\ref{sec:heisenberg} we presented a device (derived from 
Paul and Pavi\v ci\'c's device \cite{P&I}) for a photonic detecting 
of objects without an energy exchange. More precisely, there is a 
very high probability approaching 100\%\ that not even a single photon 
energy $h\nu$ will be transferred to the objects. Figuratively, 
one could call the device a ``Heisenberg microscope without a kick.''
In Sec.~\ref{sec:path} we employed the device in the \it welcher 
Weg\/ \rm detection of the atoms taking part in an interference 
experiment. Both, the Heisenberg microscope reasoning and arguments 
against a \it welcher Weg\/ \rm experiment traditionally rest on the 
Heisenberg uncertainty relations. Uncertainty relations always 
refer to the mean values of the operators and that means---even 
when the operators are projectors---statistics obtained by recording 
an interaction, i.e., by a reduction of the wave packet. 
In our \it ``energy--exchange--free microscope''\/ \rm 
measurement (Sec.~\ref{sec:heisenberg}) we do not attach 
any value to any operator in the Hilbert space description of the 
observed systems and therefore, \it no\/ \rm uncertainty relation is 
involved. As for the \it welcher Weg\/ \rm experiment 
(Sec.~\ref{sec:path}) it has recently been shown that 
``it is possible to obtain 
\it welcher Weg\/ \rm information without exposing the interfering
beam to uncontrollable scattering events...  That is to say, it is 
simply the information contained in a functioning measuring apparatus 
that changes the outcome of the experiment and not uncontrollable 
alterations of the spatial wave function, resulting from the action 
of the measuring apparatus on the system under observation.'' 
\cite{scully} There is, however, an essential difference between 
our proposal and the ones by Scully, Englert, and Walther 
\cite{scully} (microwave cavity proposal), by Sanders and Milburn 
\cite{QND} (quantum nondemolition measurement with the Kerr medium) 
and by Paul \cite{paul} (perfectly reflecting mirror proposal). In 
all of them there is slight exchange of energy which does not 
significantly disturb the spatial wave function of the system 
taking part in the interference but does disturb its phase. 
In our proposal we apparently have no exchange of energy. We say 
``apparently'' because in a future real experiment we should 
discuss the Bohrian physical process responsible for 
disappearance of the interference fringes in detail.

\bigskip 
\bigskip 
\bigskip 
\parindent=20pt
I thank Harry Paul for many discussions and suggestions. 
I acknowledge supports of the Alexander von Humboldt 
Foundation, Germany, the Max--Planck--Gesellschaft, Germany, 
and the Ministry of Science of Croatia.

\begin{figure}
\caption{Lay--out of the proposed energy--exchange--free experiment; 
(a) In the shown free round--trips the intensity of the reflected 
beam is approaching 0 for $R$ approaching 1, i.e., detector $D_r$ 
does not react; (b) However, when an absorbing object is immersed 
in the liquid (whose refractive index is the same as the one of the 
crystal in order to prevent losses of the free round--trips), 
for $R=0.999$, 99.9\%\ of the incoming beam reflect into $D_r$, 
0.0001\%\ go into $D_t$, and 0.0999\%\ hit the object.}
\label{exp} 
\end{figure}

\bigskip
\begin{figure}
\caption{Realistic values of $\eta$ for $R=0.95$ (the lowest 
curve), 0.99, 0.995, 0.997, and 0.998. The curves represent the sum 
given by Eq.~(\protect\ref{eq:sum}) as a function of the number of 
round--trips.}
\label{eff} 
\end{figure}

\begin{figure}
\caption{Proposal for a \it welcher Weg\/ \rm experiment with 
ultracold atoms. MOTIRR resonators R (see Fig.~\protect\ref{exp}), 
here shown sideways, move together with the falling atoms which sit 
in their openings. See Sec.~\ref{sec:path} for other details.}
\label{path} 
\end{figure}

\vfill\eject 

\epsffile{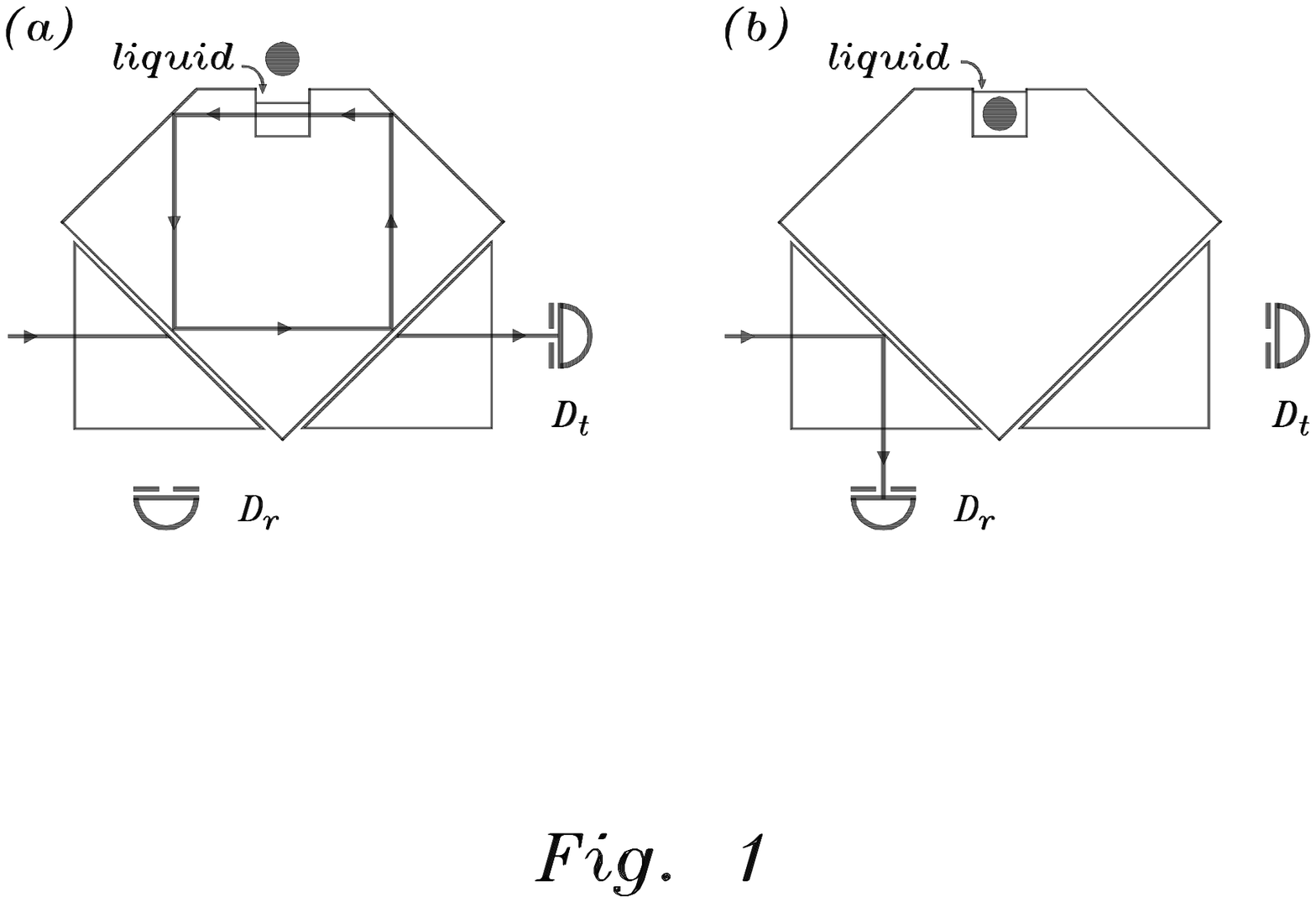}

\vfill\eject 

\epsffile{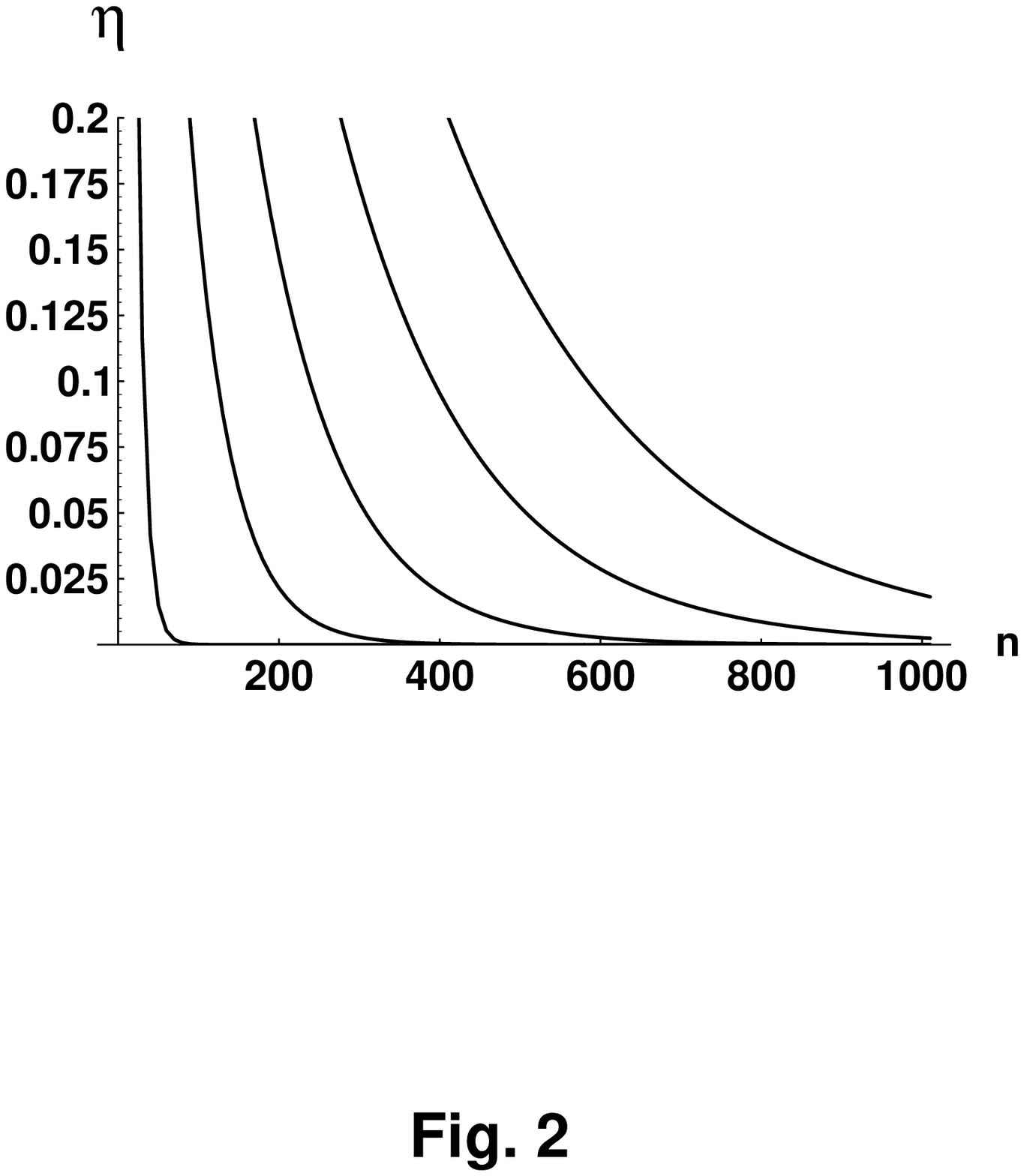}

\vfill\eject 

\epsffile{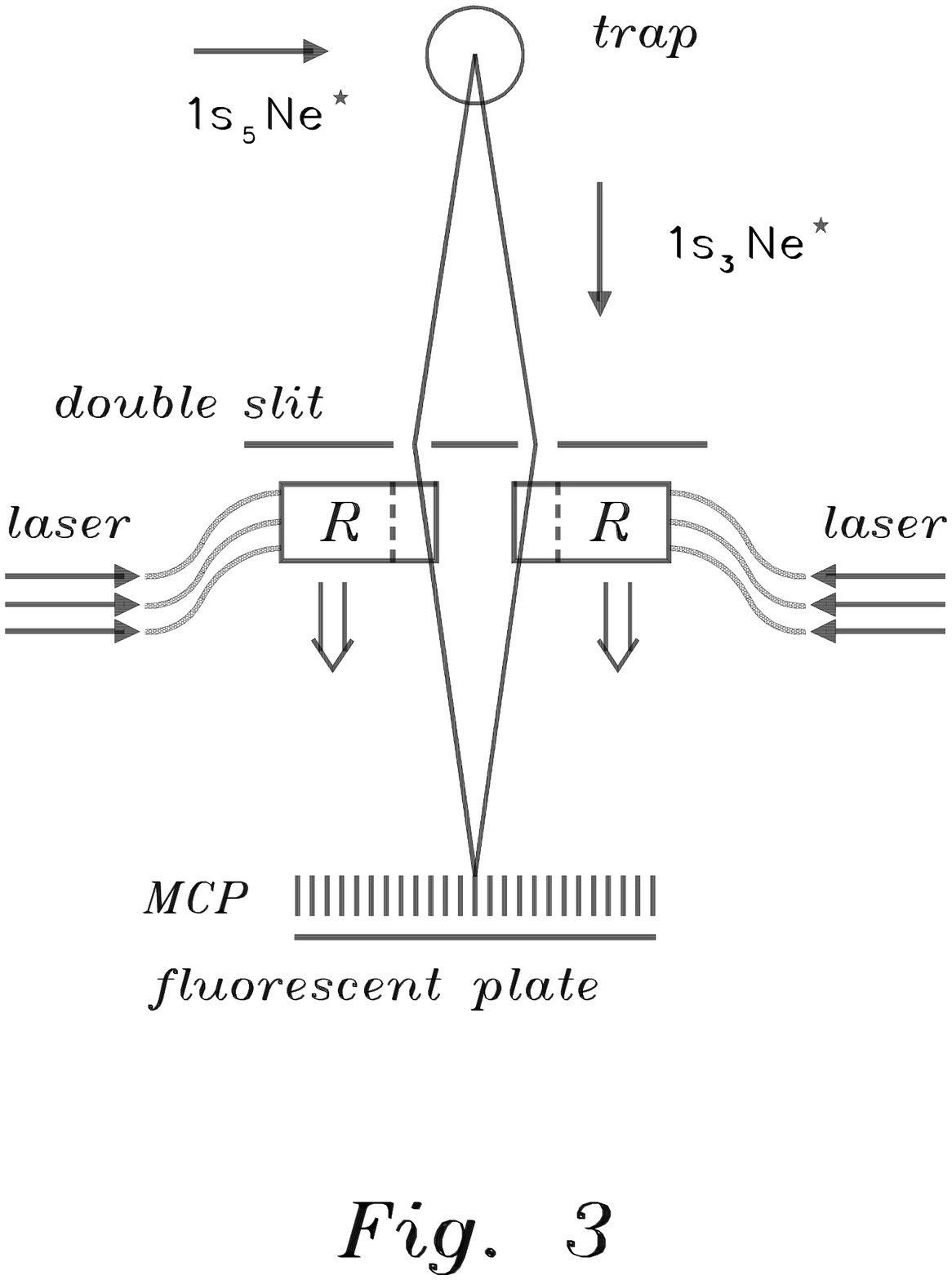}

\end{document}